\documentclass[prl,twocolumn,showpacs,floatfix]{revtex4}

\usepackage{graphicx}

\begin{document}               

\def\be{\begin{equation}}
\def\ee{\end{equation}}
\def\ba{\begin{eqnarray}}
\def\ea{\end{eqnarray}}
\def\bas{\begin{eqnarray*}}
\def\eas{\end{eqnarray*}}


\title{Ground-state properties 
of hard core bosons in one-dimensional harmonic traps}
\author{T.~Papenbrock}
\affiliation{Physics Division, 
Oak Ridge National Laboratory, Oak Ridge, TN 37831, USA}
\date{\today}

\begin{abstract}
The one-particle density matrices for hard core bosons in a
one-dimensional harmonic trap are computed numerically for systems
with up to 160 bosons. Diagonalization of the density matrix shows
that the many-body ground state is not Bose-Einstein condensed. The
ground state occupation, the amplitude of the lowest natural orbital,
and the zero momentum peak height scale as powers of the particle
number, and the corresponding exponents are related to each
other. Close to its diagonal, the density matrix for hard core bosons
is similar to the one of noninteracting fermions.
\end{abstract}


\maketitle

Bose-Einstein condensation (BEC) in dilute atomic vapors is a very
attractive field of contemporary scientific research. Of particular
interest are one-dimensional systems \cite{Druten} which are
experimentally realized in very elongated three-dimensional traps
\cite{Goerlitz}. The one-dimensional Bose gas displays a rich behavior
ranging from the weakly interacting regime \cite{tp} to the
experimentally realized Thomas-Fermi regime to a gas of impenetrable
hard core bosons \cite{Petrov,Dunjko,Girardeau01} at sufficiently low
densities and large $s$-wave scattering length. This latter case of
hard core interactions is particularly interesting since the exact
many-body ground-state wave function [See Eq.~(\ref{gs})] is known due
to a boson-fermion mapping \cite{Girardeau65}. In spite of this
knowledge, it is nontrivial to extract the ground-state properties
from this wave function. The main difficulty to overcome is the
computation of the one-particle density matrix for sufficiently large
systems. Girardeau {\it et al.} \cite{Girardeau01} used Monte Carlo
integration techniques to compute the density matrix for small systems
with up to $N=10$ bosons. More accurate methods employed by Lapeyre
{\it et al.} \cite{Lapeyre02} are restricted to up to $N=8$
particles. Analytical results were derived by Kolomeisky {\it et al.}
\cite{Kolomeisky} for the diagonal of the density matrix and by
Minguzzi {\it et al.} \cite{Minguzzi} for the tails of the momentum
distribution.  It is the purpose of this work to compute the one
particle density matrix for large systems containing more than a
hundred particles and to present analytical results and
scaling relations as well.

The ground-state wave function for $N$ hard core bosons is given by
\be
\label{gs}
\psi_B(z_1,\ldots,z_N)= C_N^{1\over 2}
\prod_{k=1}^N e^{-z_k^2/2}
\prod_{1\le i < j \le N}|z_i-z_j|
\ee
where the normalization constant is
\be
C_N=2^{N(N-1)/2}\pi^{-N/2}\left(\prod_{n=1}^N n!\right)^{-1} .
\ee
Note that the wave function (\ref{gs}) is simply the absolute value of
the ground-state wave function $\psi_F$ for $N$ noninteracting fermions 
\cite{Girardeau65},
i.e. $\psi_B=|\psi_F|$.  The one-particle density matrix is defined as
\ba
\label{densmat}
\rho_B(x,y) &=& N \int dz_1\ldots dz_{N-1}\psi_B^*(z_1,\ldots,z_{N-1},x)
\nonumber\\
&&\times\psi_B(z_1,\ldots,z_{N-1},y)
\ea
and requires the integration over $N-1$ variables. Let us rewrite 
the one-particle density matrix as
\ba 
\rho_B(x,y)&=&{2^{N-1}\over \pi^{1/2} (N-1)!} 
\exp{\left\{-{1\over 2}(x^2+y^2)\right\}}\nonumber\\
&&\times\int d\mu_{N-1} \prod_{k=1}^{N-1} |z_k-x| |z_k-y|.
\ea
Here we introduce the integration measure 
\be
\label{measure}
d\mu_{N-1}\equiv C_{N-1}\prod_{k=1}^{N-1}dz_k e^{-z_k^2}
\prod_{1\le i < j \le N-1}(z_i-z_j)^2,
\ee
which is identical to the joint probability density for eigenvalues
$z_1,\ldots,z_{N-1}$ of $(N-1)$-dimensional random matrices drawn from
the Gaussian Unitary Ensemble (GUE) \cite{Mehta,GMW}. 

This observation greatly facilitates the Monte Carlo integration since
it is easy to generate independent configurations $z_1,\ldots,z_{N-1}$
that are distributed according to the measure (\ref{measure}). In
practice, one draws a $(N-1)$-dimensional random matrix from the GUE,
i.e. the matrix is complex Hermitian, and its matrix elements are
Gaussian random variables with zero mean and variance $1/2$. Upon
diagonalization, one obtains eigenvalues $z_1,\ldots, z_{N-1}$ that are
distributed according to the measure (\ref{measure}). The integrand
$\prod_{k=1}^{N-1} |z_k-x| |z_k-y|$ is then evaluated for this
configuration, and the procedure is repeated many times. This procedure
requires an effort of $O(N^3)$ for each configuration. The
computation of the density matrix for $N=10$ bosons takes about one
hour on a PC and requires $10^6$ configurations. While this is
already a considerable improvement over previous integration
techniques \cite{Girardeau01,Lapeyre02}, it is not sufficient when much
larger systems are considered.

Following Mehta \cite{Mehta}, we express the measure~(\ref{measure}) in
terms of harmonic oscillator wave functions $\varphi_n(z)$ as 
\be
d\mu_{N-1}={dz_1\ldots dz_{N-1}\over (N-1)!}
\left({\rm
det}\left[\varphi_{k-1}(z_l)\right]_{k,l=1,\ldots,N-1}\right)^2.  
\ee
The integration can be performed since the $(N-1)$-dimensional
integral factorizes 
\ba 
\lefteqn{\int dz_1\ldots dz_{N-1} \left({\rm
det}\left[\varphi_{k-1}(z_l)|z_l-x|\right]_{k,l=1,\ldots,N-1}\right)}
\nonumber\\
&&\times\left({\rm
det}\left[\varphi_{i-1}(z_j)|z_j-y|\right]_{i,j=1,\ldots,N-1}\right)
\nonumber\\
&&= (N-1)! \,{\rm det}
\left[B_{m,n}(x,y)\right]_{m,n=0,\ldots,N-2}.  
\ea 
Here we introduce the $(N-1)$-dimensional square matrix with elements
\be
\label{Bmat}
B_{m,n}(x,y)\equiv \int\limits_{-\infty}^{\infty} dz 
                |z-x| |z-y| \varphi_m(z)\varphi_n(z).
\ee
We may thus express the one-particle density matrix of $N$ bosons as
the determinant of a $(N-1)$-dimensional matrix
\be
\label{main}
\rho_B(x,y)= {2^{N-1}e^{-(x^2+y^2)/2}\over \pi^{1/2}(N-1)!} 
{\rm det} \left[B_{m,n}(x,y)\right]_{m,n=0,\ldots,N-2}.
\ee

\begin{figure}
\includegraphics[width=0.45\textwidth]{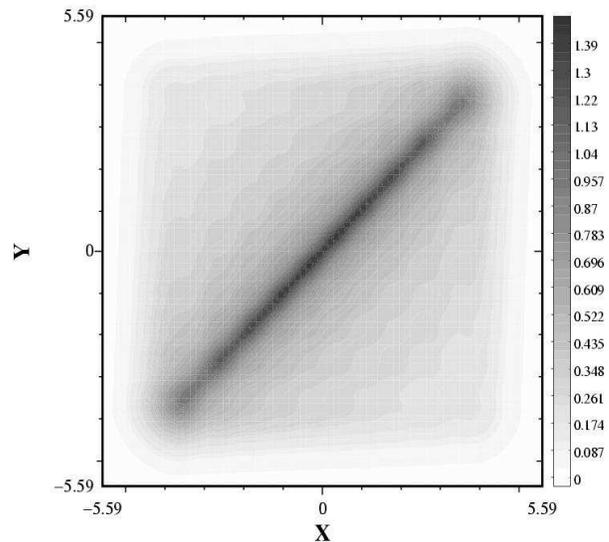}
\caption{\label{fig1}One-particle density matrix $\rho_B(x,y)$ for
$N=10$ bosons. Lightest gray indicates
almost zero amplitude, while black indicates the maximal amplitude
$\approx(2N)^{1/2}/\pi$. The width of the diagonal peak is proportional to
$N^{-1/2}$.}
\end{figure}

\begin{figure}
\includegraphics[width=0.45\textwidth]{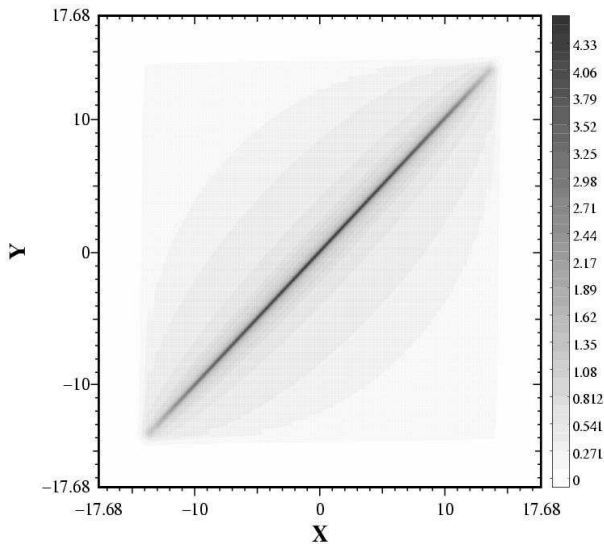}
\caption{\label{fig2}Same as Figure~\ref{fig1}, but for $N=100$ particles.}
\end{figure}

This form of the density matrix and its connection to the GUE was
previously discussed by Forrester {\it et al.} \cite{Forrester}. The
density matrix $\rho_B(x,y)$ may now be computed numerically. For a
numerical computation of Hermite polynomials, see, e.g.,
Refs.\cite{Zhang,Gautschi}. The total effort scales like $O(N^5)$, and
the calculation for a system of $N=100$ bosons requires about one day
on a PC. We computed the density matrix for systems containing up to
$N=160$ particles. Figures~\ref{fig1} and \ref{fig2} show density
plots of the density matrix $\rho_B(x,y)$ for $N=10$ and $N=100$
particles, respectively. The salient features are strong intensities
close to the diagonal $x=y$ and small off-diagonal contributions which
rapidly drop to zero at $|x|,|y| \approx R$, where
\be
\label{R}
R\equiv (2N)^{1/2}
\ee
is the radius of the density $\rho_B(x,x)$ \cite{Kolomeisky} or 
Wigner's semicircle \cite{Mehta}. Note that the width of the peak
along the diagonal decreases with increasing particle number.

Diagonalization of the density matrix yields the natural orbitals
$\phi_j(x)$ and occupation numbers $n_0\ge n_1\ge n_2\ge \ldots\ge 0$
fulfilling $\int dy \rho_B(x,y)\phi_j(y)=n_j\phi_j(x)$, i.e. 
\be
\label{spectral}
\rho_B(x,y)=\sum_j n_j\phi_j(x)\phi_j(y).
\ee
The many-body ground-state is Bose-Einstein condensed provided the
density matrix exhibits a macroscopic eigenvalue $n_0\propto N$; the
corresponding natural orbital $\phi_0(x)$ is then the wave function of
the BEC. Our data show that $n_0$ is not simply a power of $N$ for the
limited range of particle numbers considered in this work. Previous
studies on small systems suggested that $n_0\propto N^\alpha$ with
$\alpha\approx 0.59$ \cite{Girardeau01}. While this value is a good
fit for systems with up to $N=10$ particles, we find that the exponent
$\alpha$ actually decreases with increasing particle number $N$,
reaching $\alpha\approx 0.53$ for the largest particle numbers
considered in this work. Below we will present simple scaling
arguments which suggest that $\alpha=0.5$ is the expected behavior for
large particle number $N$. This behavior of harmonically trapped hard
core bosons is similar to the uniform system of hard core bosons
\cite{Lenard}.

\begin{figure}
\includegraphics[width=0.45\textwidth]{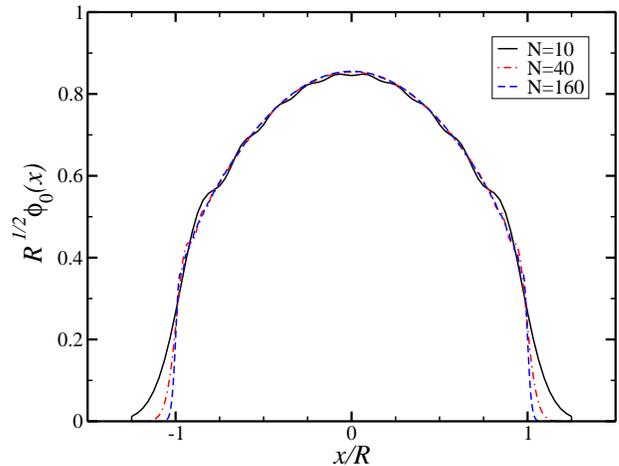}
\caption{\label{fig3}Scaled natural orbital $R^{1/2}\phi_0(x)$ for
$N=10$ (full line), $N=40$ (dashed-dotted line), and $N=160$ (dashed line) 
bosons. $R=(2N)^{1/2}$ sets the length scale in units of the oscillator 
length.  The natural orbitals
$\phi_0(x)$ are nonzero on an interval that scales like $N^{1/2}$ and
have a maximum amplitude proportional to $N^{-0.25}$.}
\end{figure}

Figure~\ref{fig3} shows the scaled natural orbitals $R^{1/2}\phi_0(x)$
for $N=10,40,160$ bosons. Note that the scaled natural orbitals
approach an $N$-independent function that depends only on the scaled
variable $x/R$ as $N$ increases. This function is nonzero only for
$|x|/R\alt 1$ and varies smoothly over this interval. We find that the
maximum amplitude of the lowest natural orbital scales like
$\phi_0(0)\propto N^\beta$ with $\beta=-0.25$. This is expected since
the natural orbital $\phi_0$ is normalized and is supported on a
domain that scales like $R\propto N^{1/2}$.

\begin{figure}
\includegraphics[width=0.45\textwidth]{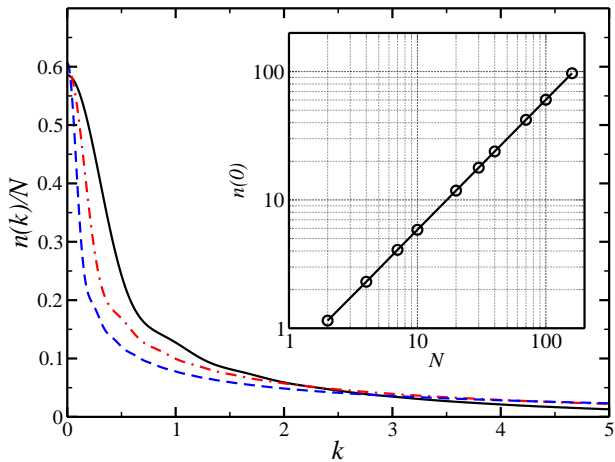}
\caption{\label{fig4}Normalized momentum distribution $n(k)/N$ for systems
of $N=10$ (full line), $N=40$ (dashed-dotted line), and $N=160$ (dashed line)
bosons. The momentum $k$ is given in units of the inverse oscillator 
length. The inset shows the zero-momentum peak $n(0)$ as a function of 
particle number $N$ in a log-log plot.}                  
\end{figure}

The momentum distribution 
\be
\label{mom}
n(k)\equiv {1\over 2\pi}\int dx dy \rho_B(x,y)e^{-ik(x-y)}
\ee
is of particular interest. Plots of the normalized momentum
distribution $n(k)/N$ are shown in Figure ~\ref{fig4} for $N=10, 40,
160$. The distributions have a pronounced peak at zero momentum and
long tails. Minguzzi {\it et al.} \cite{Minguzzi} showed that the
tails of the momentum distribution decay as $n(k)\propto k^{-4}$ for
large momenta $k$. We find that the peak height $n(0)$ is proportional
to the particle number, $n(0)\propto N^\gamma$ with $\gamma=1.0$ (see
the inset of Figure~\ref{fig4}). Thus, the system of hard core bosons
mimics the macroscopic occupation of a momentum zero state and in this
aspect resembles a uniform and noninteracting Bose system.

This finding is particularly interesting, because it allows us to 
predict the $N$-dependence of the ground-state occupation $n_0$. According
to Eq.~(\ref{mom}), the momentum peak height $n(0)$ is simply the integral 
over the density matrix which we approximate as
\ba
n(0)&\propto&
\!\!\!\!\!\!\!\!\int\limits_{\mbox{diagonal region}}
\!\!\!\!\!\!\!\!\!\!\!\!\!\!\!\! dx dy \,\rho_B(x,y)\nonumber\\ 
&+& n_0\!\!\!\!\!\!\!\!\!\!\!\!\!\!\!\!
\int\limits_{\mbox{off-diagonal domain}} 
    \!\!\!\!\!\!\!\!\!\!\!\!\!\!\!\!dx dy \,\phi_0(x)\phi_0(y).
\ea
Here we have decomposed the domain of integration into the diagonal
region over the peaked structure and the off-diagonal contribution;
the latter may be approximated in leading order by the integral over
the lowest natural orbital $\phi_0$ since all other natural orbitals
have smaller occupation number and their positive and negative
amplitudes lead to cancellations when integrated over the off-diagonal
domain. The contribution from diagonal region scales like $N^{1/2}$
and can therefore be neglected when compared with the momentum peak
$n(0)\propto N^{1.0}$. This can be seen as follows. The normalized
density along the diagonal is proportional to $N$, while the width of
the diagonal peak decreases like $N^{-1/2}$ with increasing particle
number (see below).  The remaining off-diagonal contribution scales
like $n_0 R^2 (\phi_0(0))^2=n_0 N^{0.5}$ since the domain of integration
is a square whose area scales like $R^2\propto N$.  Thus, $n_0\propto
N^\alpha$ with $\alpha=0.5$. Let us summarize the analysis of this
paragraph into the equation 
\be
\gamma=1+\alpha+2\beta
\ee 
that relates the scaling of the momentum peak height to the ground
state occupation and the amplitude of the lowest natural orbital.

The density matrix $\rho_B(x,y)$ can be understood analytically close
to its diagonal $x=y$. Kolomeisky {\it et al.} \cite{Kolomeisky}
showed that the diagonal density matrix is identical to the density
of noninteracting fermions in one-dimensional harmonic traps,
i. e. $\rho_B(x,x)=\sum_{n=0}^{N-1}\varphi(x)\varphi(x)$,
which is also identical to the level density of GUE random matrices
\cite{Mehta}. This analogy holds also in leading order as one leaves
the diagonal. To see this, we write the matrix (\ref{Bmat}) as 
\be
\label{B2}
B_{m,n}(x,y)= F_{m,n} - 2\int\limits_{\min{(x,y)}}^{\max{(x,y)}} dz
(z-x)(z-y) \varphi_m(z)\varphi_n(z), 
\ee 
with 
\bas
\label{F}
\lefteqn{F_{m,n}\equiv \int_{-\infty}^{\infty} dz 
                (z-x)(z-y) \varphi_m(z)\varphi_n(z)} \\
&&=\left(xy+n+{1\over 2}\right)\delta_m^n 
-{x+y\over \sqrt{2}}\left(\sqrt{n}\delta_m^{n-1}+\sqrt{m}\delta_{m-1}^n\right)
\\
&&+ {1\over 2} \left(\sqrt{m(n+1)}\delta_{m-1}^{n+1}
                  +\sqrt{n(m+1)}\delta_{m+1}^{n-1}\right).
\eas

The decomposition of the matrix elements $B_{m,n}$ into the
matrix elements $F_{m,n}$ and a remainder is quite useful. The
remaining integral in Eq.~(\ref{B2}) vanishes on
the diagonal $x=y$ and yields corrections proportional to $|x-y|^3$ as
one leaves the diagonal. The leading corrections proportional to
$(x-y)^2$ are already contained in the matrix elements
$F_{m,n}$. Thus, close to the diagonal, the matrix $B$ can be
approximated by the matrix $F$. This latter matrix, however,
determines the density matrix $\rho_F(x,y)$ for noninteracting
fermions in a one-dimensional harmonic trap. Indeed, adjusting the
above calculation to the case of noninteracting fermions yields
\bas
\rho_F(x,y)= {2^{N-1}e^{-(x^2+y^2)/2}\over \pi^{1/2}(N-1)!} 
{\rm det} \left[F_{m,n}(x,y)\right]_{m,n=0,\ldots,N-2}
\eas
for the fermionic density matrix.
[Note that the more familiar expression $\rho_F(x,y)=\sum_{n=0}^{N-1}
\varphi_n(x)\varphi_n(y)$ for the same density matrix can be obtained
in a direct calculation].  Thus, the density matrix for hard core
bosons is practically identical to the density matrix for
noninteracting fermions close to its diagonal. Note, finally, that the
square of the fermionic density matrix $\rho_F^2(x,y)$ is identical to
the two-level cluster function of the GUE \cite{Mehta}.  This allows
us to transfer results obtained for random matrices to the case of
noninteracting fermions. Close to the origin, we thus find for large
particle number $N\gg 1$ and $\delta \ll N^{-1/2}$
\bas
\rho_B(\delta,-\delta)\approx \rho_F(\delta,-\delta)
={1\over \pi}\sqrt{2N}\left(1-{4\over 3} N\delta^2\right).   
\eas
Thus, the width and height of the prominent peak along the diagonal of
the bosonic density matrix (see, e.g., Figures~\ref{fig1},
\ref{fig2}) scales like $N^{-1/2}$ and like $N^{1/2}$, respectively.
This behavior is confirmed by our numerical computations where we take
the width at half maximum.

In summary, we expressed the density matrix of $N$ harmonically
trapped hard core bosons as a determinant of a $(N-1)$-dimensional
symmetric matrix and performed numerical computations on systems
containing more than a hundred particles.  The density matrix is
strongly peaked along its diagonal but lacks off-diagonal long range
order. Accordingly, the ground-state is not Bose-Einstein
condensed. The ground-state occupation, the amplitude of the lowest
natural orbital, and the momentum peak height scale as powers of the
particle number, and the corresponding exponents are related to each
other by a simple equation.  The lowest natural orbital approaches a
particle number independent function when scaled properly. Analytical
results show that the density matrices of hard core boson systems and
noninteracting fermion systems are almost identical close to the diagonal.

The author acknowledges communications with P. J. Forrester. This
research used resources of the Center for Computational Sciences at
Oak Ridge National Laboratory (ORNL). ORNL is managed by UT-Battelle,
LLC for the U.S. Department of Energy under contract
DE-AC05-00OR22725.

\end{document}